

\magnification=\magstep1
\hsize 15.4true cm
\vsize 23.0true cm
\nopagenumbers
\headline={\ifnum \pageno=1 \hfil \else\hss\tenrm\folio\hss\fi}
\pageno=1

\hfill IUHET--416

~
\vskip 1.0cm

 
\noindent{\bf WHITHER HADRON SUPERSYMMETRY?}

\bigskip\bigskip
\hskip 2cm  D. B. Lichtenberg
\medskip

\hskip 2cm Physics Department 

\hskip 2cm Indiana University

\hskip 2cm Bloomington, IN 47405, USA
\bigskip
\hskip 2cm (Invited talk to be given at Orbis Scientiae, 
Fort Lauderdale, 

\hskip 2cm December 16--19, 1999. A revised version will be
published

\hskip 2cm in the proceedings.)

\bigskip \bigskip

\item{} {\bf Abstract.} A dynamically broken hadron
supersymmetry appears to exist as a consequence of QCD. 
The reasons for the supersymmetry appear most transparently
in the framework of the constituent quark model with a
diquark approximation to two quarks. Applications
of the supersymmetry have led to relations between
meson and baryon masses and to predictions that 
certain kinds of exotic hadrons should not be observed. 
I summarize the successful applications and discuss
possible future directions for this research.

\vskip 2cm

\noindent {\bf INTRODUCTION}
\bigskip
Physicists have applied the concept of supersymmetry
to a number of different areas.
To particle physicists, the  most familiar supersymmetry 
is a spontaneously broken supersymmetry between particles 
and sparticles, for which at present no experimental 
evidence exists. However, there is experimental evidence
for dynamically broken supersymmetries in the areas of
atomic physics, nuclear physics, and hadron physics.


As far as I know, the oldest of these applications
of supersymmetry was to hadron physics, discussed first by 
Miyazawa [1] in 1966. 
Almost a decade later, Catto and G\"ursey [2]
made plausible that dynamically broken hadron 
supersymmetry is a consequence of QCD. They also showed
that one consequence of the supersymmetry is that 
Regge trajectories of mesons and baryons have 
approximately the same slope.

The reason for hadron supersymmetry is most transparent
in the approximation to QCD known as the constituent
quark model. 
In this model, the reason for hadron
supersymmetry can be seen as follows: According to QCD
an antiquark belongs to a $\bar {\bf 3}$ multiplet
of color SU(3). A two-quark system, which I call a 
diquark, can be in either a {\bf 6}
or $\bar {\bf 3}$ multiplet. 
Any two constituent quarks in
a baryon must belong to the $\bar {\bf 3}$ so that the
baryon can be an overall color singlet. 
Now a meson contains a constituent quark and a constituent
antiquark. If we replace the 
antiquark (a fermion) by a $\bar {\bf 3}$ 
diquark (a boson), we make a supersymmetric transformation 
of a meson into a baryon. This transformation does not 
change the color 
configuration. Because, in first approximation, the 
QCD interaction depends only on the color configuration,
the force between the quark and diquark in a baryon should 
be approximately the same as the force between quark and
antiquark in a meson. Hence, we should be able to 
use supersymmetry to relate
the properties of baryons to the properties of mesons. 

If we replace antiquarks by diquarks in normal hadrons, 
we can obtain exotic hadrons. For example, if we replace
$Q\bar Q$ (a bar on the symbol for a particle denotes
the antiparticle) by $D \bar D$, where $Q$ is a quark and 
$D$ is a diquark, we obtain an exotic meson from a normal
one. Making use of supersymmetry, we can relate properties
of exotic hadrons to similar properties of normal ones. 
In exotic hadrons, a diquark can be either in a
$\bar {\bf 3}$ or a {\bf 6} multiplet of color. The 
interactions
of the {\bf 6} cannot be related by supersymmetry
to the interactions of an antiquark, and so we must
neglect the {\bf 6}. We justify this neglect as follows:  
When two quarks are
close together, QCD says that their Coulomb-like 
interaction is attractive in a $\bar {\bf 3}$ and 
repulsive in a {\bf 6}. It is then plausible that the
$\bar {\bf 3}$ lies lower in energy than the {\bf 6}. 
If we confine ourselves to low-mass exotics we
hope that we may safely neglect 
the contribution of color-{\bf 6} diquarks.

The difficulty with applying supersymmetry to hadrons is 
that
the supersymmetry is badly broken, or the pion and proton
would have the same mass. Miyazawa [1] was already aware
of this difficulty in 1966. Supersymmetry breaking
arises from at least three differences between a diquark
and a quark (or antiquark): 1) they have different sizes; 
2) they
have different masses; and 3) they have different spins.
We briefly discuss these differences.

1) Obviously, a diquark is not a point particle,
but neither is a constituent quark, as it consists of
a pointlike quark surrounded by a cloud
of gluons and quark antiquark pairs. I have not seen
any paper discussing how supersymmetry is broken by
size differences between quark and diquark, and I have
made no progress on this problem myself, so
I have to neglect the effects of diquark size.

2) Mass effects can be taken into account in several
ways. One particularly simple method is to relate
mass differences between mesons to mass differences
between baryons in such a way that the effects of the
diquark-quark mass difference is most likely to cancel
out. Another method is to make use of the fact 
that the quark-antiquark binding energy in mesons 
depends smoothly on the constituent quark 
masses [3]. In this method, the binding energy
of a quark with a diquark can be estimated by treating
the diquark as a fictitious antiquark with the diquark
mass.

3) There are spin-dependent forces in QCD. One way to
minimize their effect is to take appropriate averages
over spin. Another way is to assume that the 
spin-dependent interaction energy  between two quarks 
in a diquark is independent of the hadrons in which
the diquark is embedded. This assumption is not strictly
correct [4], but it is a good approximation. Then the
spin-dependent contribution to  the interaction energy
can be approximately extracted [4] from
the experimentally known masses of baryons. 
In both spin averaging and extracting
spin-dependent forces from baryons, it is assumed that the
spin-dependent force in ground-state hadrons is the
usual chromomagnetic force arising from one-gluon 
exchange [5].
This assumption has been challenged by Glozman
and Riska [6], and I have
discussed the arguments in favor of one-gluon
exchange in a talk at the last Orbis meeting [7].

I have been working on the consequences of broken hadron
supersymmetry for several years and have spoken about
it at two previous Orbis meetings [8,9]. In the present talk
I shall update the conclusions of my two earlier talks
and discuss possible directions for future work in hadron
supersymmetry. 

\bigskip\medskip
\noindent {\bf RELATIONS BETWEEN MESON AND BARYON MASSES}
\bigskip
    From here on, I will sometimes call an
antiquark a quark and an antidiquark a diquark. In this
language, for example, a meson is a two-quark state and 
an exotic meson is a four-quark or two-diquark state.

We use the notation that $Q$ denotes any quark, $q$
denotes a light $u$ or $d$ quark, and $D$ ($QQ$) denotes a
color $ \bar {\bf 3}$ diquark. Also $M$ ($Q\bar Q$) is
a normal meson, $M_E$ ($QQ\bar Q\bar Q$ or $D\bar D$) is
an exotic meson, $B$ ($QQQ$ or $QD$) is a normal baryon, 
$B_E$ ($QQQQ\bar Q$ or $DD\bar Q$) is an exotic baryon,
and $B_2$ ($QQQQQQ$ or $DDD$) a dibaryon. 

As a consequence of hadron supersymmetry, we can make the 
transformations 
$$\bar Q \rightarrow D, \quad Q \rightarrow \bar D .
\eqno(1)$$
Applying either the first or second of eqs.\ (1) one or
more times,  we obtain
$$M=Q\bar Q  \rightarrow B= QD, \eqno(2)$$ 
$$B=QD  \rightarrow M_E= \bar D D, \eqno(3)$$
$$\bar B =\bar Q \bar Q \bar Q
\rightarrow B_E =DD\bar Q, \eqno(4)$$
$$\bar B =\bar Q\bar Q \bar Q
\rightarrow B_2= DDD. \eqno(5)$$

We next consider how to take into account supersymmetry
breaking. One way to minimize the effects of
spin-dependent forces is to average over spins in such
a way that perturbatively the spin-dependent forces
cancel out. In order to do this, we must make an
assumption about the nature of these spin-dependent
forces. Following De R\'ujula et al.\ [5], we assume
that the spin-dependent forces arise from one-gluon
exchange. Then the spin averaging of ground-state
hadrons is given by the prescription of
Anselmino et al.\ [4]. One way to minimize the effects
of mass differences between quarks and diquarks is
to let one quark in the diquark be a light quark $q$.
We do this by confining ourselves (in this section)
to the transformations
$$\bar Q \rightarrow D_q =Qq, \quad
Q \rightarrow \bar D_q =\bar Q \bar q. \eqno(6)$$
We also take differences in masses such that the
effect of the extra light quark in the diquark will
tend to cancel out.

In the following, we let the symbol for a hadron
denote its mass, and we write the constituent
quarks of a hadron in parentheses following the
hadron symbol. We are led by the considerations of
the previous paragraph to consider the difference
of two meson masses: $M(\bar Q_2 q)-M(\bar Q_1 q)$. 
Applying the transformation of eq.\ (6), we get
$$M(\bar Q_2 q)-M(\bar Q_1q) = B(Q_2qq)-B(Q_1qq). \eqno(7)$$
The masses in eq.\ (7) are to be thought of as 
spin averages, i.e.
$$M(\bar qq) = (3\rho + \pi)/4, \quad M(\bar sq) = 
(3K^* + K)/4,$$
$$M(c\bar q) = (3D^* + D)/4, \quad M(\bar b q) =
(3B^* + B)/4, \eqno(8)$$
$$B(qqq) = (\Delta + N)/2, \quad B(sqq) = (2\Sigma^*
+ \Sigma + \Lambda)/4,$$
$$B(cqq) =(2\Sigma_c^* + \Sigma_c + \Lambda_c)/4, \quad
B(bqq) =(2\Sigma_b^* + \Sigma_b + \Lambda_b)/4, \eqno(9)$$
where the symbols for the mesons and baryons
are those of the 
Particle Data Group [10]. Using eqs.\ (8) and (9) in (7),
we obtain the sum rules [8]
$$(3K^* + K)/4 -(3\rho + \pi)/4 =
(2\Sigma^* + \Sigma + \Lambda)/4-
(N+\Delta)/2, \eqno(10)$$
$$(3D^* + D)/4 -(3K^* + K)/4=
(2\Sigma_c^* + \Sigma_c + \Lambda_c)/4-
(2\Sigma^* + \Sigma + \Lambda)/4, \eqno(11)$$
$$(3B^* + B)/4-(3D^* + D)/4=
(2\Sigma_b^* + \Sigma_b + \Lambda_b)/4-
(2\Sigma_c^* + \Sigma_c + \Lambda_c)/4. \eqno(12)$$
These same sum rules were obtained earlier [11] by a 
method not using hadron supersymmetry. However, 
the assumption of one-gluon exchange was needed for
averaging over spin states.

We can test the sum rules with the experimental values
of the known hadron masses [10]. The left-hand side of 
eq.\ (10)
is $182\pm 1$ MeV, while the right-hand side is
$184\pm 1$ MeV, in good agreement with experiment.
Similarly, the left-hand side of eq.\ (11) is 
$1179 \pm 1$ MeV, while the right-hand side is
$1174 \pm 1$ MeV, also in satisfactory agreement with
the data. In a 1996 talk [8], I noted that eq.\ (12)
was consistent with preliminary data on baryons
containing $b$ quarks, but the
1998 tables of the Particle Data Group [10] do not
confirm those data. Therefore, the sum rule of eq.\ (12)
remains to be tested by experiment. 

The fact that the sum rules of  eqs.\ (10) and (11) 
agree with the data constitutes evidence in support of 
spin-dependent forces arising from one-gluon exchange.
These sum rules do not follow from the spin-dependent
forces postulated by Glozman and Riska [6]. In their 
work, the spin-dependent forces in baryons containing
only light quarks arise from pseudoscalar meson 
exchanges. However, I don't see how the same mechanism
can apply to mesons or to baryons containing heavy 
quarks. If I would need two or three different 
mechanisms to account for the spin-dependent forces in
hadrons (or a linear combination of them), then
I would not know how to obtain sum rules.

\bigskip\medskip
\noindent{\bf EXOTIC HADRONS}
\bigskip

We do not need to restrict ourselves to spin-averaged
hadron masses or to diquarks containing at least one
light quark, as we can explicitly take into account
mass and spin effects. I discussed this problem at
a previous Orbis [9], and so will only briefly review
the method. 

We start with the spin-averaged hadron masses, but
include spin effects explicitly at a later stage.
We assign constituent masses to the quarks such 
that the binding energy of a quark and antiquark in a
meson is a smooth
function of the reduced mass of the two constituents [3].
We can use this ``meson curve'' to read off
the binding energy of a fictious hadron made of
a fictious quark and antiquark of any given masses.

We consider a spin-averaged baryon made of a quark and a 
diquark, treating
the diquark as a fictious antiquark. Our first guess
for the diquark  mass is that it equals 
the sum of its  two constituent quark masses. 
We obtain the reduced mass of the quark and diquark and
read off the binding energy from  the meson curve. 
We add this
binding energy to the masses of the quark and diquark
to obtain a calculated spin-averaged baryon mass. 
In general, this mass does not equal the experimental
mass of a baryon, averaged over spin. However, by 
repeatedly adjusting the mass 
of the diquark, we can obtain the correct spin-averaged 
baryon mass. We are thus able to obtain the 
spin-averaged diquark
masses for constituent quarks of any flavors. 

Next we obtain diquark properties from observed baryon 
masses rather than from spin-averaged masses.
We extract the spin-dependent interaction energies of 
two quarks in a diquark from the observed baryon masses
[4]. Adding these terms to the spin-averaged diquark
mass, we obtain the masses of spin-one and spin-zero
diquarks. 

We are now ready to calculate the masses of 
ground-state exotic hadrons.
We first consider exotic mesons containing at least one
diquark of spin zero. In such mesons, there are no 
spin-dependent forces between the diquarks. Therefore,
we only have to calculate the reduced mass of the 
constituents and add the binding energy from the meson 
curve to the diquark masses in order to obtain the exotic
meson mass.
(If both diquarks have spin one, there are additional
spin-dependent forces, but their effects can be 
calculated.)

The results of these calculations is that 
diquark-antidiquark exotic mesons
have sufficiently large masses to  decay rapidly into two 
normal mesons. Because we expect production cross sections 
to be small and decay widths large,  it is unlikely
that such exotic mesons will be observed. A
possible exception is that an exotic meson containing a $bb$
diquark might be stable against strong decay, but its
production cross section will be extremely small. 
Our conclusion is in agreement with the fact that
no exotic mesons composed of a diquark and antidiquark have
yet been seen. 

The same method can be applied to exotic baryons and
to dibaryons. However, there is  the complication that,
except in the limit of point-like diquarks, the Pauli
principle is not strictly satisfied for quarks in different
diquarks. The results are similar to the results for mesons:
exotic baryons and dibaryons (other than the deuteron)
are not likely to be observed. Again, this conclusion
is in agreement with observations to date. 


\bigskip\medskip
\noindent{\bf THE FUTURE}
\bigskip

The predictions of the previous sections follow
from broken hadron supersymmetry plus spin-dependent
forces arising from one-gluon exchange. It is gratifying
that we have not obtained any predictions in serious
disagreement with experiments done so far, but it is
disappointing that our model says that diquark exotics
will probably not be observed.

Although enough has been established so far to give
me confidence that hadron supersymmetry is a useful
concept, open questions remain to be answered.
Among them are:

(1) A diquark may be almost as large as the hadron that 
contains it. How do we correct for the non-negligible 
size of a diquark? 

(2) Is there any way to take into account the contribution
from color-sextet diquarks to exotic hadrons?

(3) If the spin-dependent forces in some hadrons
are not given by
one-gluon exchange but rather by the mechanism
of Glozman and Riska, how do the results change? Are the
changes large enough to destroy the good agreement with
experiment?

(4) How can we take the Pauli principle into account
in exotic baryons and dibaryons?  

(5) Exotic hadrons containing diquarks can mix with other
hadrons having the same quantum numbers. For example,
quantum numbers permitting, a diquark-antidiquark meson
can mix with normal mesons, hybrids, and glueballs. 
Can we take this mixing into account? 

(6) Are there any other useful predictions to be
obtained from broken hadron supersymmetry?

In conclusion, if physicists can successfully tackle the
preceding open questions, hadron supersymmetry will rest
on a much sounder foundation than it does now. 
However, if answers are
not forthcoming, it may be time for physicists to 
store in their minds that broken hadron supersymmetry
exists and go on to other topics. 

\bigskip\medskip
\noindent{\bf ACKNOWLEDGMENTS}
\bigskip

Some of this work was done with Enrico Predazzi and 
Renato Roncaglia. I should like to thank Florea Stancu
for helpful discussions.


\bigskip\bigskip
\noindent {\bf REFERENCES}
\medskip
\font \eightrm=cmr9
\par
\everypar{\parindent=0pt \hangafter=1\hangindent=2 pc}

\eightrm
\par
\noindent 1. H. Miyazawa,
Baryon number changing currents,
{\it Prog.\ Theor.\ Phys.}\ 36:1266  (1966).


2. S. Catto  and F. G\"ursey,
Algebraic treatment of effective supersymmetry,
{\it Nuovo Cimento} 86:201 (1985).

3. R. Roncaglia, A.R. Dzierba, D.B. Lichtenberg,
and E. Predazzi, Predicting the masses of heavy hadrons
without an explicit Hamilatonian,
{\it Phys.\ Rev.\ D} 51:1248 (1995).

4. M. Anselmino, D.B. Lichtenberg, and E. Predazzi,
Quark color-hyperfine interactions in baryons,
{\it Z.\ Phys.\ C} 48:605 (1990).

5. A. De R\'ujula, H. Georgi, and S. L. Glashow,
Hadron masses in a gauge theory,
{\it Phys.\ Rev.\ D} 12:147 (1975).

6. L.Ya Glozman  and D.O. Riska, The spectrum of the
nucleons and the strange hyperons and chiral dynamics,
{\it Phys.\ Rep.}\ 268:263 (1996).

7.  D. B. Lichtenberg, Spin-dependent forces between
quarks in hadrons, Orbis Scientiae, Fort Lauderdale
(Dec.\ 18--21, 1998), B.N. Kursunoglu et al., eds.
Plenum Press, New York (1999). 

8. D. B. Lichtenberg,
Hadron Supersymmetry and Relations Between
Meson and Baryon Masses, Orbis Scientiae, Miami Beach
(Jan.\ 25--28, 1996), in Neutrino Mass, Dark Matter,
Gravitational Waves, Monopole Condensation and Light
Cone Quantization, B. Kursunoglu, S.L. Mintz, and A.
Perlmutter, eds., Plenum Press, New York (1996), pp.\
319--322.
 
9. D.B. Lichtenberg, Exotic Hadrons,
Orbis Scientiae, Miami Beach (Jan.\ 23--26, 1997),
in {\it High Energy Physics and
Cosmology: Celebrating the Impact of 25 Gables Conferences}
B.N. Kursunoglu, S.L. Mintz, and A. Perlmutter, eds.
Plenum Press, New York (1997), pp.\  59--65.

10.  Particle Data Group: C. Caso et al., Review of 
Particle Physics, {\it Euro.\ Phys.\ J. C} 3:1 (1998).

11. D.B. Lichtenberg and R. Roncaglia,   
New formulas relating the masses of some baryons and mesons,
{\it Phys.\ Lett. B} 358:106 (1995).

\bye